# The Impact of a STEM-based Entrepreneurship Program on the Entrepreneurial Intention of Secondary School Female Students

Mojtaba Shahin[*1], Olivia Ilic[1], Chris Gonsalvez[1], Jon Whittle[2]

[1]Monash University, Melbourne, Australia, [2]CSIRO's Data61, Australia

mojtaba.shahin@monash.edu, olivia.ilic@monash.edu, chris.gonsalvez@monash.edu, jon.whittle@data61.csiro.au

## Abstract

Despite dedicated effort and research in the last two decades, the entrepreneurship field is still limited by little evidence-based knowledge of the impacts of entrepreneurship programs on the entrepreneurial intention of students in pre-university levels of study. Further, gender equity continues to be an issue in the entrepreneurial sector, particularly in STEM-focused entrepreneurship. In this context, this study was designed to explore the effects of a one-day female-focused STEM-based entrepreneurship program (for brevity, we call it the OzGirlsEntrepreneurship program) on the entrepreneurial intention of secondary school female students. The study collected data from two surveys completed by 193 secondary school female students, aged 14-16 years, who participated in the OzGirlsEntrepreneurship program. This program encouraged girls to develop and implement creative computational solutions to socially relevant problems, with an Internet of Things (IoT) component using the micro:bit device. The findings reveal that a key factor in the development of entrepreneurial attitudes in young female students is associated with soft-skills development, particularly in the areas of creative thinking, risk-taking, problem-solving, and leadership development. The importance of meaningful human connections, including positive role modelling and peer to peer learning were also important factors in fostering entrepreneurial intent. With these factors in mind, our findings highlight that the OzGirlsEntrepreneurship program substantially increased the entrepreneurial intention of secondary school female students. In addition, this study offers actionable implications and recommendations to develop and deliver entrepreneurship education programs for secondary school level students.

**Keywords**: Entrepreneurial intention, Entrepreneurship education, learning, inspiration, secondary school female students, female entrepreneurs

## 1 Introduction

Entrepreneurs impact the economic and social growth of countries through creating new jobs, fostering innovation in markets, and boosting competition (KritiKoS, 2014; Van Praag & Versloot, 2007). Realising the promising benefits of entrepreneurship requires entrepreneurship education programs that increase entrepreneurship awareness and cultivate entrepreneurial attitude, mindset, behaviour, and intention among young people (O'Connor, 2013; Oosterbeek, van Praag, & Ijsselstein, 2010; José C Sánchez, 2013). This view has motivated many policymakers, governments, and academic institutions to design and organise different types of entrepreneurship programs, for example, introducing compulsory or elective entrepreneurship courses in university curricula and/or organising entrepreneurship programs as an extra-curricular activity for university graduates (Farnell, Heder, & Ljubić, 2015; Fellnhofer, 2019; Martin, McNally, & Kay, 2013; Ozaralli & Rivenburgh, 2016; Pittaway & Cope, 2007). The growing interest in entrepreneurship programs can be attributed to the assumption: "entrepreneurs can be made" (Erikson, 2003; José C Sánchez, 2013).

---

[*] Corresponding author at: Faculty of Information Technology, Monash University, Melbourne, Australia. E-mail address: mojtaba.shahin@monash.edu (Mojtaba Shahin)

The research around entrepreneurship education has been mostly conducted in the context of university education, where the subjects are graduate or postgraduate students (Fayolle & Gailly, 2015; Karimi, Biemans, Lans, Chizari, & Mulder, 2016; Nabi, Walmsley, Liñán, Akhtar, & Neame, 2018; Pittaway & Cope, 2007). Very limited attention has been devoted to non-university individuals and, in particular, to secondary school students (Elert, Andersson, & Wennberg, 2015; Peterman & Kennedy, 2003; Rodrigues, Dinis, do Paço, Ferreira, & Raposo, 2012; José C Sánchez, 2013). Hence, the impact of entrepreneurship programs on secondary school students (particularly on their entrepreneurial intents) has remained mostly untested (José C Sánchez, 2013). Researchers highlight the critical role of secondary schools in innovation and entrepreneurship (Fagerberg & Srholec, 2009; José C Sánchez, 2013). They refer to the childhood and teenage years as the best time to learn entrepreneurship skills and develop an entrepreneurial intention (Filion, 1994; Gasse, 1985). Skills, knowledge, and attributes that strengthen entrepreneurial intent include learning and inspiration in the areas of opportunity identification, creative problem-solving, positive role-modelling, and teamwork, to name a few (Johannisson, 1991; Karimi et al., 2016; Nabi, Holden, & Walmsley, 2010). Furthermore, the development of intention towards entrepreneurship during childhood and adolescence is highlighted as an important factor to promote a persistent interest in creating commercial or social ventures later (Degeorge & Fayolle, 2008; Dyer, 1995; Elert et al., 2015; Low, Yoon, Roberts, & Rounds, 2005; Nguyen, Do, Vu, Dang, & Nguyen, 2019).

In considering the role of entrepreneurship in education, it is critical to acknowledge the diversity, or lack thereof, of the actors within the student populous. For the most part, male entrepreneurs are seen to make up a larger proportion of the student population in entrepreneurship courses (Bosma & Levie, 2010; Cochran, 2019; Dabic, Daim, Bayraktaroglu, Novak, & Basic, 2012; Daim, Dabic, & Bayraktaroglu, 2016; Menzies & Tatroff, 2006). Female entrepreneurs however are said to play a significant role in entrepreneurship, with diversity contributing to greater capacity for innovation (Mascarenhas & Vander Veer, 2014) and economic growth across the globe (Ahl, 2006; Allen, Langowitz, & Minniti, 2007; Shinnar, Giacomin, & Janssen, 2012). While the last decade has witnessed a growing trend in female entrepreneurship, gender equity continues to be an issue in the entrepreneurial sector (Bosma & Kelley, 2019; Shinnar et al., 2012). For example, the rate of working women who were self-employed was 9.6%, while this share was 16.9% for men in the European Union in 2018 (Union, 2019). Some researchers (Díaz-García & Jiménez-Moreno, 2010; Westhead & Solesvik, 2016; Wilson, Kickul, & Marlino, 2007) also found that women are less, if not significantly less, interested in entrepreneurial activities and engaging in the entrepreneurial process. A set of environmental and cultural constraints contribute to this imbalance, for example, a lack of high-profile female-owned firms, gender stereotypes (e.g., gender stereotypes may lead women to perceive their entrepreneurship skills as lower than that of men), and the higher level of difficulties that women face in accessing and obtaining resources to start a new business or building an effective entrepreneurial network (Halabisky, 2018; Kariv, 2013; Kuschel, Ettl, Díaz-García, & Alsos, 2020; Marlow & Swail, 2014; Welter, 2011; Westhead & Solesvik, 2016). This imbalance gets worse in fields that are dominated by men, such as STEM (Science, Technology, Engineering, and Mathematics) (Poggesi, Mari, De Vita, & Foss, 2020). Failure to have an increased number of females in the STEM and entrepreneurial sectors can inevitably lead to a decrease in innovative and productive capacity and overall economic competitiveness (Ashcraft & Blithe, 2010; Simard, 2008; Voyles, Haller, & Fossum, 2007). Female-only entrepreneurship programs that promote entrepreneurial competencies such as opportunity identification are a promising education model to enhance the intention of women to set up businesses in STEM fields (Armuña, Ramos, Juan, Feijóo, & Arenal, 2020; Boddington & Barakat, 2018). Furthermore, compared to mixed-gender educations programs, this type of education program helps women feel safer, more comfortable, and self-confident (Boddington & Barakat, 2018). It is argued that more research should be conducted to increase awareness and interest in STEM-focused entrepreneurship for women (Elliott, Mavriplis, & Anis, 2020; Kuschel et al., 2020; Poggesi et al., 2020).

With this in mind, this paper focuses on outlining the key findings of a one-day female-focused STEM-based entrepreneurship program (i.e., for brevity, we call it the OzGirlsEntrepreneurship program) for secondary school aged female students (henceforth referred to as "students") alongside actionable implications for educators, program designers, and mentors who operate within the growing ecology of entrepreneurship education. It analyses the effect of the program on the entrepreneurial intention of the students as well as the factors (e.g., prior exposure to entrepreneurship) influencing this effect. Focusing efforts on female-focused STEM-based entrepreneurship programs has the capacity to increase diversity in the workforce, keeping labour markets competitive and ensuring girls receive an equitable and opportunity filled future (Barker & Aspray, 2006; Elliott et al., 2020; Margolis, Goode, & Bernier, 2011; Miliszewska & Sztendur, 2010; Teague, 2002). Emphasising technology-oriented innovation promotes diversity in the workforce and adds breadth to the makeup of teams, increasing their problem-solving, creative and innovative capacity (Ashcraft & Blithe, 2010; Barker & Aspray, 2006; Catherine Ashcraft, 2012; Papastergiou, 2008).

The OzGirlsEntrepreneurship program had 203 Year 10 female student participants from 44 secondary schools in the state of Victoria in Australia. Year 10 in Australia is the final year of compulsory secondary education, in which students' ages range from 14 to 16. The students were grouped into 52 teams of 3-4 members. They were asked to develop and implement creative computational solutions with an Internet of Things (IoT) component to socially relevant problems using the micro:bit device[1], in the context of problem-based learning strategy (Hmelo-Silver, 2004). Data collection was conducted through two surveys completed by students.

Our findings are: (1) The OzGirlsEntrepreneurship program increased the intention of secondary school female students to become an entrepreneur at a significant level. (2) The students who indicated greater entrepreneurial inspiration and learning (Souitaris, Zerbinati, & Al-Laham, 2007) from the OzGirlsEntrepreneurship program showed a significantly higher attitude towards entrepreneurship. (3) The students who indicated greater entrepreneurial inspiration and learning from the OzGirlsEntrepreneurship program showed a higher entrepreneurial intention. This increase is at a significant level when entrepreneurial attitude plays a mediating role. (4) There is a positive and statistically strong link between secondary female students' entrepreneurial attitude and entrepreneurial intention. It is believed that three key factors are at play here: (1) The importance of creative problem-solving in developing an entrepreneurial attitude, (2) Meaningful human connections to build entrepreneurial intent and (3) An alignment to real-world problems. These factors are highlighted with the intent to provide a contextual framework for the ongoing design, delivery, and development of future entrepreneurship programs conducted within the education sector. In doing so, it is hoped that the positive results will encourage educators to deliver these types of programs, giving more young females the opportunity to experience and enjoy the benefits of entrepreneurship education.

We develop our hypotheses in the context of existing literature in Section 2. Our research method is explained in Section 3. We detail our findings in Section 4, and in Section 5 we discuss the key findings, report on the limitations, and provide recommendations for research and practice.

## 2 Background and Hypotheses

### 2.1 Entrepreneurial Attitude and Intention

From a psychological perspective, entrepreneurship (action) is a planned, intentional behaviour, which can be strongly expressed and predicted by entrepreneurial intention (Bird, 1988; Katz & Gartner, 1988; Krueger, Reilly, & Carsrud, 2000). This has resulted in widespread interest among entrepreneurship practitioners and researchers in understanding what factors influence entrepreneurial intention **(EI)** (Fayolle & Liñán, 2014; Maresch, Harms, Kailer, & Wimmer-Wurm, 2016). Moriano, Gorgievski, Laguna, Stephan, and Zarafshani (2012) define EI as "the conscious state of mind that precedes action and directs attention toward entrepreneurial behaviours such as starting a new business and becoming an entrepreneur" (p. 165). EI can be influenced directly by the constructs of intention-based models, as well as indirectly by exogenous influences (also known as situational and individual factors) (Krueger et al., 2000; Souitaris et al., 2007). Theory of planned behaviour (TPB) developed by Ajzen (1991) is a widely operationalised entrepreneurship intention-based model. According to Ajzen's model, "attitude towards the behaviour (entrepreneurship)", "subjective norms", and "perceived behavioural control", as three attitudinal constructs of intention, can predict EI. Researchers have applied Ajzen's model to different contexts to predict individuals' EIs, broadly confirming the predictive power of this model (Engle et al., 2010; Karimi et al., 2016; Krueger et al., 2000; Liñán & Chen, 2009; Nguyen et al., 2019; Ozaralli & Rivenburgh, 2016). In this study, we only focus on the prediction of EI through one of the constructs (i.e., "attitude towards entrepreneurship") in Ajzen's model. The "attitude towards entrepreneurship" construct (i.e., entrepreneurial attitude **(EA)**) refers to "the degree to which a person has a favourable or unfavourable evaluation or appraisal of becoming self-employed" (Ajzen, 1991) (p. 188). This implies that having the intention to launch a business venture generally requires first forming an attitude towards making a new business venture. Many studies found that attitude is a predictor of intention (Lortie & Castogiovanni, 2015). For example, Krueger et al. (2000) found the EIs of senior university students are significantly predicted by their attitudes towards entrepreneurship. An empirical study by Sun, Lo, Liang, and Wong (2017) in Hong Kong showed that engineering students'

---

[1] https://microbit.org/

attitudes towards entrepreneurship and their EIs are positively and statistically associated. A vast majority of the current entrepreneurship research has been studying the link between entrepreneurial attitude **(EA)** and the EIs of university students and nascent entrepreneurs. Still, there exists little research concerning EI in teenage years (Nguyen et al., 2019). Researchers have emphasised that early development of EI is of great importance, as it may promote a persistent interest in creating an entrepreneurial venture later (Degeorge & Fayolle, 2008; Dyer, 1995; Nguyen et al., 2019). We, therefore, hypothesise:

**Hypothesis 1 (H1)**. Students' entrepreneurial attitude is positively associated with their entrepreneurial intention.

## 2.2 Entrepreneurship Education and Intention

Research has empirically investigated the impacts of exogenous factors such as education, gender, skills, cultural factors, work experience, and financial support on EI (Adekiya & Ibrahim, 2016; Boyd & Vozikis, 1994; Krueger et al., 2000; Shapero & Sokol, 1982; Souitaris et al., 2007). The exogenous factors indirectly change EI by affecting other factors, e.g., changing attitudes towards self-employment (Krueger et al., 2000). Among exogenous influences, there is less dedicated research investigating how different types of entrepreneurship education programs **(EEPs)** change students' EI. This has recently led to a growing number of researchers empirically investigating this potential relationship (Barba-Sánchez & Atienza-Sahuquillo, 2018; Karimi et al., 2016; Nabi et al., 2018; Souitaris et al., 2007). EEPs, which may be organised in the form of different types of initiatives such as curricular or extracurricular activities, and compulsory or elective programs (Arranz, Ubierna, Arroyabe, Perez, & Fdez. de Arroyabe, 2017), aim at developing, nurturing, and improving entrepreneurial intention, inspiration, mindset, and knowledge (Ozaralli & Rivenburgh, 2016). The results linking entrepreneurship education **(EE)** and EI are somewhat controversial and are sometimes inconsistent (Karimi et al., 2016; Ozaralli & Rivenburgh, 2016). Some scholars have confirmed that EE has a strong positive influence on EI (Adekiya & Ibrahim, 2016; Souitaris et al., 2007), while other researchers observed that while this relationship is positive, the strength of the relationship is not statistically strong (Karimi et al., 2016; Liñán & Fayolle, 2015; Martin et al., 2013; Nowiński, Haddoud, Lančarič, Egerová, & Czeglédi, 2019). Furthermore, some research indicates that EI is negatively associated with EE (Nabi et al., 2018; Oosterbeek et al., 2010). For instance, Oosterbeek et al. (2010) observed that the intent to be an entrepreneur among a vocational college' students in the Netherlands were significantly reduced after participating in a compulsory entrepreneurship program. These contradictory findings may stem, for example, from pedagogy (i.e., how entrepreneurship is taught) and content (Cui, Sun, & Bell, 2019; Moberg, 2014). Among the growing body of research attempting to explore the link between EE and EI, very little work has examined this link in the pre-university setting (Cárcamo-Solís, Arroyo-López, Alvarez-Castañón, & García-López, 2017; Rodrigues et al., 2012; José C Sánchez, 2013) or first-year undergraduate students (Nabi et al., 2018), particularly in the context of pre-university female students (Ferri, Ginesti, Spanò, & Zampella, 2018; Poggesi et al., 2020). Entrepreneurship researchers (Barba-Sánchez & Atienza-Sahuquillo, 2018; Nabi et al., 2018) argue that more research is required to empirically investigate the potential effects of EE on young people's EIs. This matter becomes much more important when we know that entrepreneurship can be encouraged and developed (Cui et al., 2019; Erikson, 2003; Karimi et al., 2016), and some of the key entrepreneurship competencies such as social competence emerge and are developed during childhood (Lans, Blok, & Gulikers, 2015). We, therefore, hypothesise:

**Hypothesis 2 (H2)**. The OzGirlsEntrepreneurship program, as an entrepreneurship education program, increases the entrepreneurial intention of students.

## 2.3 Entrepreneurial Inspiration and Intention

Earlier, we discussed that an entrepreneurship program as a whole might have both positive and negative impacts on EI. Souitaris et al. (2007) assert that the increased intention towards entrepreneurship that a participant achieves during an entrepreneurship program may also stem from the benefits that the participant gains from the entrepreneurship program. They argue that *entrepreneurial inspiration* is one of the main gains of entrepreneurship programs. A "program-derived entrepreneurial inspiration" (henceforth referred to as "entrepreneurial inspiration" for simplicity) is defined as: "a change of hearts (emotion) and minds (motivation) evoked by events or inputs from the program and directed towards considering becoming an entrepreneur" (Souitaris et al., 2007) (p. 573). Moreover, entrepreneurship programs encompass triggers such as, observing and interacting with entrepreneurial role models, which can inspire individuals considering and pursuing a career as an entrepreneur (Souitaris et al., 2007; Thrash & Elliot, 2003). A few studies have explored the link between inspiration and entrepreneurial attitude and intention. The studies (Nabi et al., 2018; Souitaris et al., 2007) found that there is a positive link between inspiration and EI. Nabi et al. (2018) also qualitatively observed that first-year undergraduate students attending an entrepreneurship program indicated higher entrepreneurial inspiration compared to those who did not. In work by Cui et al. (2019), it was found that inspiration is a mediator between EE and entrepreneurial mindset, which means inspiration accounts for *all* or *part* of the relationship between EE and entrepreneurial mindset (Baron & Kenny, 1986). A set of studies have analysed the effect of inspiration on entrepreneurial attitude, intention, and behaviour through *inspiring* role models (Bosma, Hessels, Schutjens, Praag, & Verheul, 2012; Krueger et al., 2000; Nowiński & Haddoud, 2019; Van Auken, Fry, & Stephens, 2006; Van Auken, Stephens, Fry, & Silva, 2006). Generally, they have shown that entrepreneurial role models can inspire and motivate students to pursue an entrepreneurship career and established a positive link between role models and entrepreneurial attitude and intention. While the studies mentioned above confirm that role models are able to boost EA and EI, Nowiński and Haddoud (2019) observed that the influence of role models on EI would be at the highest level, when considered in conjunction with positive attitudes towards self-employment and entrepreneurial self-efficacies. All this suggests that inspiration *does* affect entrepreneurial attitude and intention. However, no study has been conducted at the secondary education level. We hypothesise:

**Hypothesis 3 (H3)**. The students who indicate greater entrepreneurial inspiration from the OzGirlsEntrepreneurship program will show a higher attitude towards entrepreneurship.

**Hypothesis 4 (H4)**. The students who indicate greater entrepreneurial inspiration from the OzGirlsEntrepreneurship program will show a higher entrepreneurial intention.

**Hypothesis 5 (H5)**. Students' entrepreneurial attitude mediates the relationship between entrepreneurial inspiration and entrepreneurial intention.

## 2.4 Entrepreneurial Learning and Intention

Entrepreneurial learning can be another benefit of an entrepreneurship program and points to the knowledge about entrepreneurship that a learner develops from an entrepreneurship program (Souitaris et al., 2007). Johannisson (1991) classifies the knowledge about entrepreneurship into five levels: "why entrepreneurs act" (i.e., known-why), "what needs to be done" (i.e., known-what), "how to do it" (i.e., known-how), "who should we know" (i.e., known-who), and "when to act" (i.e., known-when). Research focusing on university students suggests entrepreneurial learning influences one's attitude and intentions towards entrepreneurship (Beliaeva, Laskovaia, & Shirokova, 2017; Nabi et al., 2018; Souitaris et al., 2007; Sun et al., 2017; F. Zhang, Wei, Sun, & Tung, 2019). Whilst Souitaris et al. (2007) found this influence to be positive, but not significant, others (Nabi et al., 2018; F. Zhang et al., 2019) observed that the EIs of university students are positively associated with the perception of their learning from an entrepreneurship program at a significant level. F. Zhang et al. (2019) have also investigated the mediating factors in this relationship, revealing that "attitude towards entrepreneurship", "subjective norms", and "perceived behavioural control" significantly mediate the link between entrepreneurial learning and EI. To best of our knowledge, no study has been conducted at the secondary education level in this regard. Hence, we hypothesise:

**Hypothesis 6 (H6)**. The students who indicate greater entrepreneurial learning from the OzGirlsEntrepreneurship program will show a higher attitude towards entrepreneurship.

**Hypothesis 7 (H7)**. The students who indicate greater entrepreneurial learning from the OzGirlsEntrepreneurship program will show a higher entrepreneurial intention.

**Hypothesis 8 (H8)**. Students' entrepreneurial attitude mediates the relationship between entrepreneurial learning and entrepreneurial intention.

Fig. 1 outlines the hypotheses mentioned above and the ways that entrepreneurial learning and inspiration affect entrepreneurial attitude and, ultimately, entrepreneurial intent.

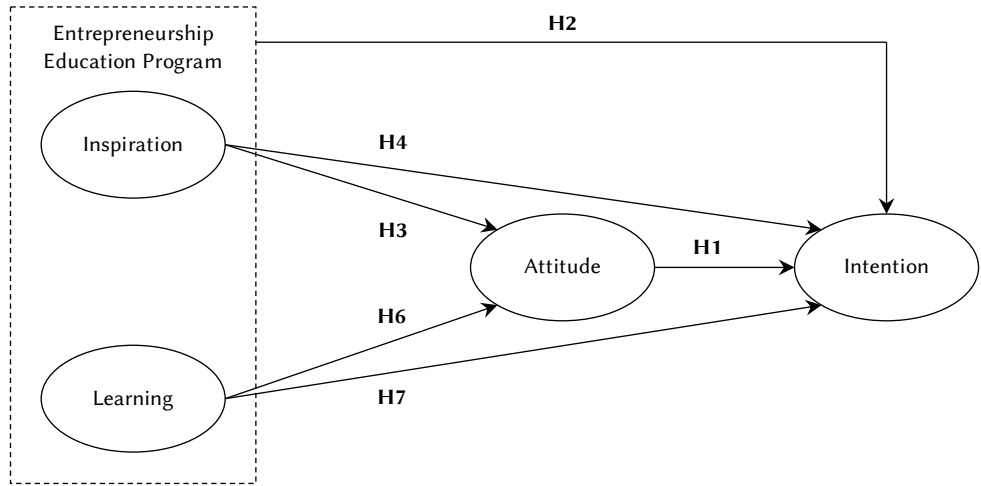
Fig. 1 Research Model (note: **H5** and **H8** are not presented for the reason of simplification)

## 3 Methodology

The OzGirlsEntrepreneurship program was the first stage of a three-stage education program (i.e., Women in STEM and Entrepreneurship program). The Women in STEM and Entrepreneurship (WISE[2]) program pursues three high-level goals: (1) developing and nurturing a technology-focused entrepreneurial intention amongst young women in Australia; (2) increasing awareness and participation of young women in STEM; and (3) understanding how young women develop and perceive computational thinking. Specifically, the WISE program targeted Year 10 female students (14-16 years old) who were studying at secondary schools in the state of Victoria in Australia in 2019 (henceforth referred to as "students"). Students attend secondary schools for six years in the Australian education system, and attendance is mandatory until age 17. Year 10 is the start of senior secondary school, and students' ages vary from 14 to 16 at this year level. All activities and data collection in the WISE program were conducted under ethics approval obtained from our university. Students and their parents/guardians were asked to complete relevant consent forms before participating in the WISE program. In this paper, we only report and analyse the data collected from the OzGirlsEntrepreneurship program as the first stage of the WISE program. The OzGirlsEntrepreneurship program included three full-day workshops held on 11th and 18th May, and 1st June 2019. In total, we hosted 203 students from 44 secondary schools. Each student participated in only one workshop (i.e., each workshop hosted approximately 70 students).

### 3.1 OzGirlsEntrepreneurship Program as a STEM-based Entrepreneurship Education Program

The OzGirlsEntrepreneurship program was designed based on the entrepreneurship program design guidelines proposed by Souitaris et al. (2007) and considering the age of our participants. The OzGirlsEntrepreneurship program also included features and activities that better suited STEM-based entrepreneurship programs. First, it was a team-based entrepreneurship program, as we grouped 203 students into 52 teams of 3-4 students. It is highlighted that STEM-focused entrepreneurship is mostly team-based (Kuschel et al., 2020; Neumeyer & Santos, 2020). Second, we added the mentoring aspect to the OzGirlsEntrepreneurship program to help students form an identity, which in turn fostering women's intention to become an entrepreneur in STEM (Elliott et al., 2020). During the OzGirlsEntrepreneurship workshops, each team was mentored by a female university student, who had recently completed or was currently studying a STEM-related degree at our university (henceforth referred to as "mentor"). Before the OzGirlsEntrepreneurship program started, we organised a one-day training workshop to instruct the 31 mentors about the objectives and logistics of the OzGirlsEntrepreneurship program and the activities that they had to carry out during the OzGirlsEntrepreneurship program. We assigned each mentor to only one student team per OzGirlsEntrepreneurship workshop, with some mentors guiding teams across multiple workshops. Mentors were selected based on their mentoring or volunteering experience and interests. Third, it heavily focused on identifying and evaluating potential opportunities because improving this type of entrepreneurial competency

---

[2]https://www.monash.edu/it/wise/

is expected to foster entrepreneurial intention in potential STEM entrepreneurs (Kuschel et al., 2020). The OzGirlsEntrepreneurship program was composed of the following components:

**Icebreaker**: First, hundreds of different pictures were shown to students, and each student had to select two pictures that represented something about themselves. Then, students were grouped into bigger groups of 9 to 15 students (we referred to this bigger group as a tribe), and were asked to introduce themselves and their school in one minute, and their picture choices.

**Role Model**: The second component was focused on the entrepreneurial process, with an emphasis on how to identify, recognise, and create business opportunities (Nabi et al., 2018; Souitaris et al., 2007). First, a female entrepreneur gave a talk about her entrepreneurial journey. Then, another female entrepreneur discussed how to develop an entrepreneurial solution for a real problem and the required steps to plan, design, and evaluate entrepreneurial solutions. Finally, the third female entrepreneur talked about the role of women in business and technology and introduced three successful women CEOs.

**Ideation and Solution**: The third component of the OzGirlsEntrepreneurship program included identifying and developing opportunities. Students were asked to identify personally and socially relevant problems and possible entrepreneurial solutions to them. In the first step, students were asked to brainstorm real problems that were important to their personal life and/or society. The members of each team then collaboratively analysed the identified problems and chose one problem for the next solution step. For example, one team decided to help hospitals prioritise patients, and another team focused on people who have trouble communicating their food allergies when they travel overseas. After identifying a problem, teams had to complete the following problem statement: "*We believe that [the identified problem] ... is a problem for [who] ... because [reason] ...*".

The next step included proposing possible solutions with an Internet of Things (IoT) component to solve the identified problem. Once the solution was identified, teams were encouraged to fill out the following solution statement: "*We could solve this problem by [solution idea] ... and could demonstrate this on the micro:bit by [prototype idea] ...*". Teams used the micro:bit device to prototype and realise their entrepreneurial ideas. The micro:bit is a RAM-base programmable IoT device designed by the BBC (i.e., released officially in 2016) to educate young learners about the principles of computer science and programming (Videnovik, Zdravevski, Lameski, & Trajkovik, 2018). Other devices such as Arduino[3] and Raspberry Pi[4] could be used instead of the micro:bit device in this study. We chose the micro:bit device because the micro:bit is specially designed for education and has lower barriers compared with its competitors (e.g., there is no need to run a full operating system in the micro:bit) for inexperienced users to understand and develop IoT-based systems (Devine et al., 2019).

Following the problem-based learning approach (Hmelo-Silver, 2004), students were responsible for identifying the knowledge deficiencies and strengths that had in solving the identified problem and seeking new knowledge to solve the problem (i.e., known as self-directed learning). It should be noted that students had the freedom (i.e., and also were encouraged) to seek help and feedback from other teams to evaluate and validate their entrepreneurial ideas, assumptions, and solutions. Furthermore, mentors (and role models) *only* guided and facilitated the problem-solving process of their teams – they were not directly involved.

**Pitching**: In the last component of the OzGirlsEntrepreneurship program, teams had to prepare a 3-minute presentation about their entrepreneurial product. During the presentation, teams were asked to talk about the problem that they solved, the potential customers/users of their product, the solution that they proposed, how their solution incorporated IoT, the proof that people would want their product, and the steps that they would take to commercialise their product in future.

### 3.2 Data Collection and Measures

Data collection was carried out through two surveys. The surveys were hosted on Google Forms. The first survey, pre-workshop survey, was conducted at the beginning of each OzGirlsEntrepreneurship workshop and consisted of 33 questions. At the end of each OzGirlsEntrepreneurship workshop, we distributed a post-workshop survey with 40 questions. Only one question in the post-workshop survey was optional. Here, we only discuss those questions from the pre- and post-workshop surveys which were used in this paper. The questions are presented in the Appendix of this paper (See Table 10 and Table 11 in Appendix):

---

[3] https://www.arduino.cc
[4] https://www.raspberrypi.org

- **Demographics**: Five questions were asked in the pre-workshop survey to collect demographic information about students, including academic performance, school type, prior entrepreneurship experience, businessperson in the immediate family, and entrepreneur in the immediate family. These questions have been widely used in the literature (Fayolle & Gailly, 2015; F. Zhang et al., 2019).

- **Entrepreneurial Inspiration**: Given each EEP is unique in terms of pedagogy and context, there is no established scale in the literature to measure entrepreneurial inspiration derived from entrepreneurship programs (Souitaris et al., 2007). We had to develop our own scale based on the experiences, activities, events, and triggers in the OzGirlsEntrepreneurship program, which might have inspired students to consider a career in entrepreneurship. Our scale consisted of 9 items, essentially reflecting the main activities of the OzGirlsEntrepreneurship program. A five-point Likert scale ("*very negative*=1" to "*very positive*=5") was adopted in the post-workshop survey to assess the impact of each item on students' mindset towards pursuing a career in entrepreneurship. Our items were (1) *Seeing role models working in entrepreneurship*, (2) *Working with my mentor*, (3) *Working with my peers*, (4) *Learning new ways of thinking*, (5) *Practising market research*, (6) *Delivering a pitch*, (7) *Connecting technology to a real-world idea*, (8) *Hands-on coding*, and (9) *Getting feedback on my ideas.*

- **Entrepreneurial Intention (EI)**: We measured the EI of students at two timeslots: at the beginning of the OzGirlsEntrepreneurship workshops (*time1*) and the end of the OzGirlsEntrepreneurship workshops (*time2*). In both timeslots, we chose to use a single-item indicator to measure the EI of students. Considering the age of students and the short timeframe (one day) of the OzGirlsEntrepreneurship program, this decision is expected to have reduced the burden on the students and kept the survey length reasonable for young females (Bowling, 2005). Furthermore, Deutskens, De Ruyter, Wetzels, and Oosterveld (2004) found that shorter questionnaires lead to a higher response rate. Other entrepreneurship researchers also used single-item questions to measure EI, e.g., (Barba-Sánchez & Atienza-Sahuquillo, 2018; Kolvereid & Isaksen, 2006; Krueger et al., 2000; Wilson et al., 2007; Y. Zhang, Duysters, & Cloodt, 2014). Single-item indicators have also been successfully used in other fields, such as marketing, e.g., (Bergkvist & Rossiter, 2007; Petrescu, 2013) and health sciences, e.g., (Ahmad, Jhajj, Stewart, Burghardt, & Bierman, 2014) to measure constructs (e.g., self-rated mental health). Similar to (Wilson et al., 2007), we sought students' level of agreement or disagreement with the statement, "*I would like to start my own business one day*" to measure their EI. This statement is one of the six items suggested by Liñán and Chen (2009) to measure EI. Furthermore, self-employment is used in economic analysis as a proxy EI (Halabisky, 2018).

- **Entrepreneurial Attitude (EA)**: We used one of five items proposed by Liñán and Chen (2009) to measure the EA of students. We asked students to rate the statement, "*A career as an entrepreneur is attractive to me*" on a five-point Likert scale ("*strongly disagree*=1" to "*strongly agree*=5"). Krueger et al. (2000) also used the same single-item indicator to measure EA. In marketing research, Bergkvist and Rossiter (2007) showed that single-item indicators could validly measure "*attitude toward the advertisement"* and "*brand attitude*" as two widely used constructs.

- **Entrepreneurial Learning**: To assess the extent to which students learned the actions required to start an entrepreneurship career (i.e., known-what) from the OzGirlsEntrepreneurship program, we used a single-item indicator ("*The OzGirlsEntrepreneurship program has increased my understanding of the actions that someone has to take in order to start an entrepreneurial career*"). This statement is one of the five scales proposed by (Souitaris et al., 2007) (rated from "*strongly disagree*=1" to "*strongly agree*=5").

### 3.3 Participants

We recruited our participants from 44 secondary schools in Victoria state, Australia. In total, 203 Year 10 female students registered for and attended the OzGirlsEntrepreneurship program. Students had the freedom to decide whether they wanted to complete the surveys. Despite this fact, 197 out of 203 students completed the pre-workshop survey (i.e., response rate: 97%) and the same students answered the post-workshop survey (i.e., response rate: 97%). It should be noted that 4 responses from each survey were discarded as we found them to be invalid (e.g., when we found the students who provided the same answer to the majority of the surveys' questions) (Meade & Craig, 2012). Finally, 193 valid responses were collected for each survey. Fig. 2 shows that 69 out of 193 (35.7%) students had previously engaged in entrepreneurship-related training before the OzGirlsEntrepreneurship program, while 124 students (64.2%) did not. Of the total 193 students, 90 (46.6%) had at least one person in their immediate family who owned or managed a company or a business. As shown in Fig. 2, only 34 students out of 193 (17.6%) students indicated that one of their immediate family members was an entrepreneur. Students from Government schools accounted for 43.5% of our participants, while 54 students (27.9%) came from Catholic schools, and 55 (28.4%) were from Independent schools. Most of the students (80.3%) indicated that their overall academic performance in class was either in the range of 80%-89.99% (86, 44.5%) or 90%-100% (69, 35.7%).

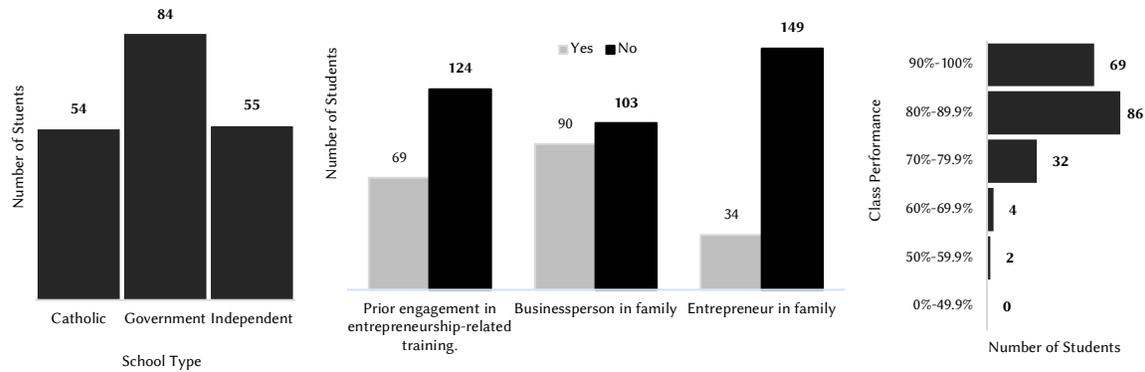

Fig. 2 Demographics

### 3.4 Data Analysis

We utilised the following statistical approaches to assess our model and hypotheses: (a) Partial least squares structural equation modelling (PLS-SEM) was adopted to appraise our model (Joe Hair, Ringle, & Sarstedt, 2011; Wold, 1982). Another method that could be used is covariance-based SEM (CB-SEM) (Joe Hair et al., 2011; Joseph Hair, Risher, Sarstedt, & Ringle, 2019; J. F. Hair, Ringle, & Sarstedt, 2012; Jöreskog, 1970). However, the accuracy and robustness of CB-SEM results highly depend on some assumptions (e.g., the sample size should be large enough – at least 200 responses), which need to be met strictly. We decided to use PLS-SEM because (1) our sample was 193 respondents; (2) our data violated the normality condition; and (3) ultimately, we aimed to evaluate to what extent entrepreneurial inspiration, learning, and attitude can predict the intentions of secondary female students towards self-deployment. (b) We applied the Wilcoxon signed-rank test (McDonald, 2009) to understand the direction and level of influence of the OzGirlsEntrepreneurship program on students' EI (i.e., hypothesis **H2**). (c) We assessed the impact of demographics on our model by using a multi-group analysis (MGA) approach in PLS-SEM (Sarstedt, Henseler, & Ringle, 2011). (d) Finally, the Mann-Whitney U test (Howell, 2012) was applied to measure the effects of demographics on students' EA and EI. We analysed our data by using SmartPLS 3 software and IBM SPSS Statistics 26 software.

## 4 Results

The literature (Chin, 2010; Joe Hair et al., 2011; Joseph Hair et al., 2019) recommends that two steps be followed to evaluate, interpret, and present PLS-SEM results. The first step is usually called "evaluation of the measurement model". It involves examining a list of criteria to appraise the reliability and validity of each construct's indicators (i.e., items) in a given model (See Section 4.1). Once the criteria are satisfied, researchers should assess the structural portion of their model, which links constructs (See Section 4.2).

### 4.1 Evaluation of the Measurement Model

**Internal consistency reliability** measures the level of consistency among the indicators (i.e., items) of a construct (Chin, 2010; Joseph Hair et al., 2019; Sharma & Stol, 2020). Our model consists of four reflective constructs; however, only *inspiration* is a multi-item construct. To check the reliability of the *inspiration* construct, we calculated Cronbach's alpha and composite reliability (CR) (See Table 1). The Cronbach's alpha $_{inspiration}$ (0.843) surpasses the cut-off of 0.70, and the CR value (0.508) is acceptable, as it is between 0.7 and 0.90 (Joe Hair et al., 2011).

Table 1. Reliability and Validity of the *inspiration* construct

| Construct | Cronbach's alpha | CR | AVE |
| --- | --- | --- | --- |
| Inspiration | 0.843 | 0.878 | 0.508 |

**Convergent validity** shows how strongly the items of a construct are positively correlated (Chin, 2010; Joseph Hair et al., 2019; Sharma & Stol, 2020). We used two techniques to examine the correlation among the items of the *inspiration* construct as the only multi-item construct in our model: "average variance extracted" (AVE) and outer loading. Any value higher than 0.50 for AVE is satisfactory (Joe Hair et al., 2011). As illustrated in Table 1, the AVE value for the *inspiration* construct (0.508) is higher than that threshold. Table 2 presents the outer loadings of the items of the *inspiration* construct. There is strong evidence in the literature (Hair Jr, Hult, Ringle, & Sarstedt, 2016; Sharma & Stol, 2020) suggesting to remove items with loadings smaller than 0.40 and retain those with loadings greater than 0.70. If removing an item with loading between 0.40 and 0.70 does not significantly improve the AVE, the item should be retained. In our model, we removed two items (i.e., ins_8 and ins_9) with loadings less than 0.70 to have increased the AVE value of the *inspiration* construct from 0.479 to 0.508. Although three items (i.e., ins_1, ins_2, and ins_3) still have loadings slightly less than 0.70, as shown in Table 2, we found removing those items could not improve the AVE. Therefore, we decided to keep those items.

Table 2. Mean, standard deviation, and cross loading of items

| Item | Item Description | Mean | SD | Attitude | Intention | Inspiration | Learning |
|---|---|---|---|---|---|---|---|
| att_1 | "A career as an entrepreneur is attractive to me" | 3.927 | 0.830 | **1.000** | 0.698 | 0.381 | 0.313 |
| int_1 | "I would like to start my own business one day" | 3.772 | 0.821 | 0.698 | **1.000** | 0.353 | 0.314 |
| ler_1 | "My understanding of the actions that someone has to take to start an entrepreneurial career has increased" | 4.627 | 0.535 | 0.313 | 0.314 | 0.289 | **1.000** |
| ins_1 | "Seeing role models working in entrepreneurship" | 4.446 | 0.682 | 0.249 | 0.228 | **0.657** | 0.300 |
| ins_2 | "Working with my mentor" | 4.477 | 0.742 | 0.107 | 0.110 | **0.634** | 0.083 |
| ins_3 | "Working with my peers" | 4.518 | 0.720 | 0.210 | 0.138 | **0.628** | 0.139 |
| ins_4 | "Learning new ways of thinking" | 4.399 | 0.669 | 0.267 | 0.289 | **0.781** | 0.242 |
| ins_5 | "Practising market research" | 4.187 | 0.766 | 0.372 | 0.348 | **0.771** | 0.170 |
| ins_6 | "Delivering a pitch" | 4.197 | 0.877 | 0.311 | 0.271 | **0.720** | 0.190 |
| ins_7 | "Connecting technology to a real-world idea" | 4.446 | 0.719 | 0.254 | 0.243 | **0.779** | 0.271 |

**Discriminant validity** shows the degree of the distinctiveness of four constructs (i.e., attitude, intention, inspiration, and learning) in our model (Joe Hair et al., 2011). Following guidelines for PLS-SEM, we assessed discriminant validity by using three approaches: "cross-loading analysis", "Fornell-Larcker criterion" (Fornell & Larcker, 1981), and "Heterotrait Monotrait (HTMT) ratio" (Henseler, Ringle, & Sarstedt, 2015). No cross-loading issue occurred because all bold numbers in Table 2 are higher than the numbers in their corresponding rows. Moreover, the difference between the highest outer loading in each row and the other loadings in that row is well above the threshold of 0.10 (Ali, Ali, Badghish, & Baazeem, 2018; Gefen & Straub, 2005). The "Fornell–Larcker criterion" was fulfilled because the square root of each construct's AVE surpasses the correlations among constructs (See Table 3) (Joe Hair et al., 2011).

Table 3. Correlations among Attitude, Intention, Inspiration, and Learning and square roots of the AVE (on diagonal)

| Construct | Attitude | Intention | Inspiration | Learning |
|---|---|---|---|---|
| Attitude | **1.000** | | | |
| Intention | 0.698 | **1.000** | | |
| Inspiration | 0.381 | 0.353 | **0.713** | |
| Learning | 0.313 | 0.314 | 0.289 | **1.000** |

While Henseler et al. (2015) state the cut-off of 0.90 for the HTMT ratio, Hair Jr et al. (2016) suggest that the value of the HTMT ratio needs to be less than 0.85 to avoid any problem in discriminant validity. Table 4 indicates that all values obtained for the HTMT ratios are smaller than the cut-off of 0.85.

Table 4. Heterotrait Monotrait (HTMT) ratios

| Construct | Attitude | Intention | Inspiration | Learning |
|---|---|---|---|---|
| Attitude | | | | |
| Intention | 0.698 | | | |
| Inspiration | 0.384 | 0.353 | | |
| Learning | 0.313 | 0.314 | 0.303 | |

## 4.2 Evaluation of the Structural Model

### 4.2.1 Assessing Collinearity

We checked the independency of the predictor constructs in our model by assessing its collinearity (Joe Hair et al., 2011). PLS-SEM suggests calculating the "variance inflation factor" (VIF) for this purpose. It is generally accepted that VIF less than 5 ensures the nonexistence of collinearity (Joe Hair et al., 2011). Still, Joseph Hair et al. (2019) believe that a more conservative threshold (i.e., VIF < 3) should be considered as there might be collinearity issues for VIFs between 3 and 5. Our model does not have any collinearity issues, as all VIFs were between 1.091 and 1.234.

### 4.2.2 Path Coefficients and Significance

*Hypothesis H1*. Fig. 3 and Table 6 show that EA was positively and substantially associated with the EI of students (*t-statistics=12.932, p-value=0.000*). Therefore, we can accept **H1**.

*Hypothesis H2*. We measured students' EIs at the start of the OzGirlsEntrepreneurship workshops (*time1*) and the end of the OzGirlsEntrepreneurship workshops (*time2*) (See Table 2). Applying the Wilcoxon signed-rank test (McDonald, 2009) shows that the OzGirlsEntrepreneurship program significantly increased the intentions of students towards entrepreneurship (time1: *mean=3.37*; time2: *mean:3.85*; *p-value=0.000*, *r=0.413*) (B. H. Cohen, 2008). Therefore, **H2** is accepted.

Table 5. The entrepreneurial intentions measured at the start of (time1) and at the end of (time2) the OzGirlsEntrepreneurship program

| | time1 Mean (SD) | time2 Mean (SD) | p-value | effect size |
|---|---|---|---|---|
| Entrepreneurial Intention | 3.37 (0.992) | 3.77 (0.823) | **0.000*** | 0.413 |

* Significant at p-value < 0.001 (Wilcoxon signed-rank test)

*Hypothesis H3*. According to the results of the bootstraps, **H3** is supported, indicating that the entrepreneurial inspirations obtained from the OzGirlsEntrepreneurship program positively and strongly related to the attitudes of female students towards entrepreneurship (*t-statistics=3.414, p-value=0.001*).

*Hypothesis H4*. Fig. 3 shows that there was a positive but not significant correlation between entrepreneurial inspiration and EI (*t-statistics=1.848, p = 0.065*). Hence, we are unable to accept **H4**.

*Hypothesis H5*. We evaluated the mediation effect of EA on the link between entrepreneurial inspiration and EI by comparing the *indirect* paths and the *direct* paths (Nitzl, Roldan, & Cepeda, 2016; Sharma & Stol, 2020; Zhao, Lynch Jr, & Chen, 2010). As illustrated in Table 6, the *indirect* relationship between entrepreneurial inspiration and EI mediated by the attitude towards entrepreneurship was statistically significant (*t-statistics=3.427, p-value=0.001*). This mediation effect is interpreted as a "*full mediation*" effect because the *direct* link between entrepreneurial inspiration and EI was insignificant (i.e., **H4** is rejected). Therefore, we accept **H5**.

*Hypothesis H6*. The results show that the students' perceived learning from the OzGirlsEntrepreneurship program positively and significantly correlated with their attitudes towards entrepreneurship (*t-statistics=2.714, p-vale=0.007*). Hence, **H6** is accepted.

*Hypothesis H7*. As evident by the path coefficient, entrepreneurial learning had a positive but insignificant relationship with the intentions of students towards entrepreneurship (*t-statistics=1.328, p-value=0.184*). Therefore, **H7** cannot be accepted.

*Hypothesis H8*. We employed a similar approach as the one we used to evaluate *Hypothesis 5* to analyse the role of EA as a mediation factor within the relation between entrepreneurial learning and EI. As shown in Table 6, EA *fully* mediated the association between entrepreneurial learning and EI (*t-statistics=2.522, p-value=0.012*). Hence, **H8** is accepted.

Table 6. The results of PLS-SEM

| | Coefficient | Sample Mean | Std.Dev | T-Statistics | 95% CI | P Values |
|---|---|---|---|---|---|---|
| H1: Attitude → Intention | 0.638 | 0.637 | 0.049 | 12.932 | (0.537, 0.728) | **0.000*** |
| H3: Inspiration → Attitude | 0.317 | 0.343 | 0.093 | 3.414 | (0.148, 0.498) | **0.001*** |
| H4. Inspiration → Intention | 0.084 | 0.091 | 0.045 | 1.848 | (-0.016, 0.165) | 0.065 |
| H5: Inspiration → Attitude → Intention | 0.202 | 0.218 | 0.059 | 3.427 | (0.096, 0.315) | **0.001*** |
| H6: Learning → Attitude | 0.221 | 0.205 | 0.082 | 2.714 | (0.061, 0.370) | **0.007** |
| H7: Learning → Intention | 0.090 | 0.084 | 0.068 | 1.328 | (-0.034, 0.231) | 0.184 |
| H8: Learning → Attitude → Intention | 0.141 | 0.132 | 0.056 | 2.522 | (0.039, 0.257) | **0.012** |

* Significant at p-value < 0.05, ** Significant at p-value < 0.01, and *** Significant at p-value < 0.001

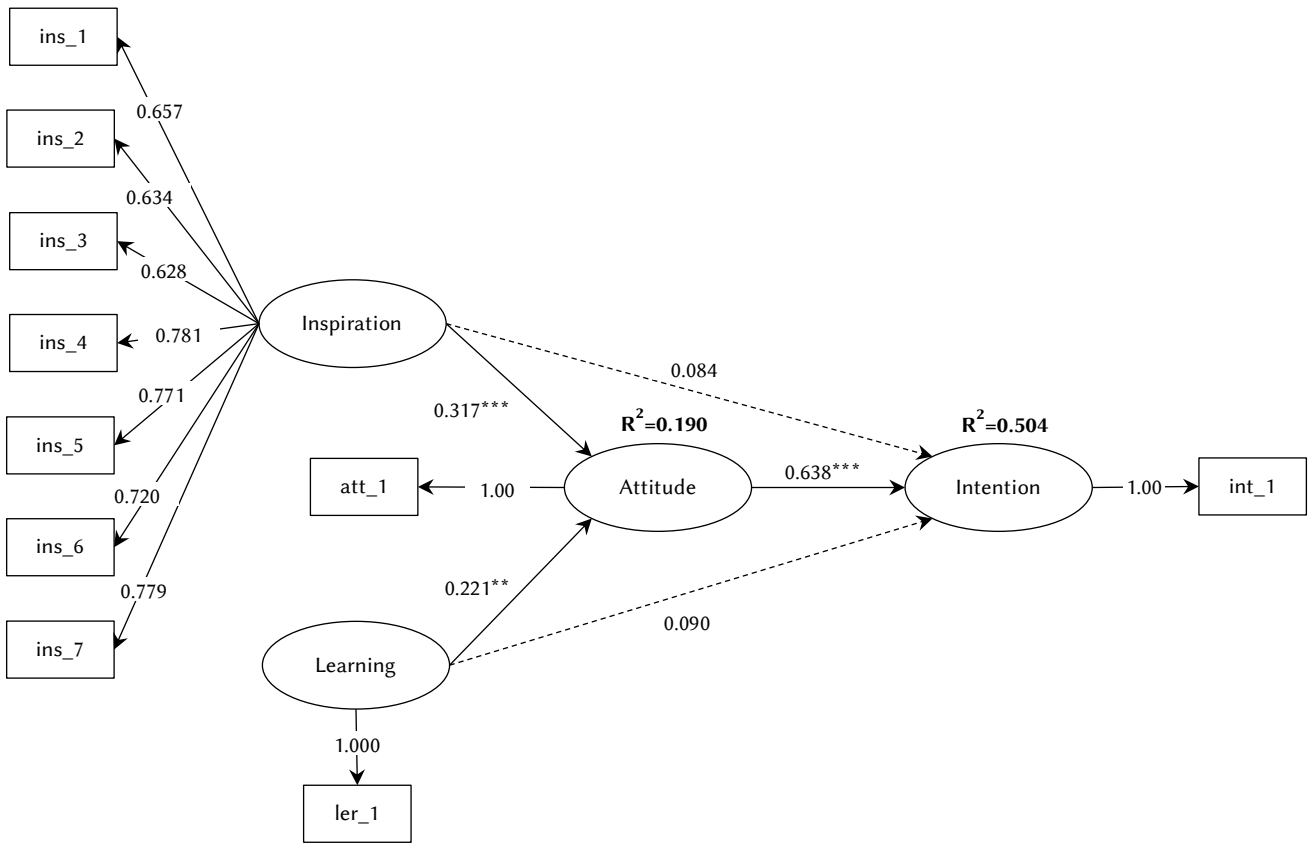

Fig. 3 Outer loadings and path coefficients (*: p-value < 0.05, **: p-value < 0.01, and ***: p-value < 0.001). Note: Dashed lines indicate the relations that are not statistically significant (Sharma & Stol, 2020).

### 4.2.3 Coefficient of Determination and Effect Sizes

The next step includes measuring the $R^2$ values of the *intention* and *attitude* constructs to examine our model's predictability (Chin, 2010). The value of $R^2$ can be between 0 and 1; however, its interpretation has been a source of controversy among scholars. Chin (2010) interprets the $R^2$ values of "0.19", "0.33", and "0.67" as "*weak*", "*moderate*", and "*substantial*". According to the guidelines developed in (Joe Hair et al., 2011; Joseph Hair et al., 2019), the values of "0.25", "0.50", and "0.75" for $R^2$ in the marketing research to be deemed "*weak*", "*moderate*", and "*substantial*" respectively. As illustrated in Table 7, the $R^2_{intention}$ is 0.504 (i.e., this signifies that *learning*, *inspiration*, and *attitude* together explain 50.4% of the variance in the *intention* construct), and the $R^2_{attitude}$ is 0.190 (i.e., this signifies that *learning* and *inspiration* jointly contribute to 19% of the variance in the *attitude* construct). Following (Joe Hair et al., 2011; Joseph Hair et al., 2019), we interpret $R^2_{intention}$ as "*moderate*" and $R^2_{attitude}$ as "*weak*".

Table 7. Coefficient of Determination

| Construct | $R^2$ | $Q^2$ |
|---|---|---|
| Intention | 0.504 | 0.474 |
| Attitude | 0.190 | 0.152 |

In the next step, we calculated the $f^2$ effect size to show the effect of removing a particular construct on the $R^2$ values of the *intention* and *attitude* constructs (Barclay, Donalds, & Osei-Bryson, 2018; Chin, 2010). J. Cohen (1988) recommends that the $f^2$ values of "0.02", "0.15", and "0.35" show "*small*", "*medium*", and "*large*" effect respectively. J. Cohen (1988) also highlighted that the $f^2$ value below 0.02 should be considered as "*no effect*". Table 8 summarises the $f^2$ values of the hypotheses in our model.

Finally, we were interested in understanding whether our model has predictive relevance (Joseph Hair et al., 2019). To this end, we calculated the Stone–Geisser's $Q^2$ value (Geisser, 1974; Stone, 1974) for the *intention* and *attitude* constructs by using the blindfolding technique (Tenenhaus, Vinzi, Chatelin, & Lauro, 2005). Both $Q^2_{intention}$ (0.474) and $Q^2_{attitude}$ (0.152) are more than 0, which means that our model has the ability to predict *attitude* and *intention* constructs (See Table 7).

Table 8. Effect sizes

| Construct | $f^2$ | Effect |
|---|---|---|
| H1: Attitude → Intention | 0.665 | Large |
| H3: Inspiration → Attitude | 0.114 | Small |
| H4: Inspiration → Intention | 0.012 | No effect |
| H6: Learning → Attitude | 0.055 | Small |
| H7: Learning → Intention | 0.014 | No effect |

### 4.2.4 Moderating Factors

We used a multi-group analysis (MGA) approach in PLS-SEM to assess the impact of different demographics on our model (Sarstedt et al., 2011). MGA can be interpreted as a special form of moderator analysis (Henseler & Chin, 2010; Henseler & Fassott, 2010). We grouped the participants based on three variables: (1) *prior entrepreneurship experience* (i.e., does a participant have prior entrepreneurship experience?); (2) *businessperson in immediate family* (i.e., does a participant have any businessperson in her immediate family?); and (3) *entrepreneur in immediate family* (i.e., does a participant have any entrepreneur in her immediate family?). Each variable is treated as a moderating factor in our model. Therefore, we had two groups for each of the variables: *Yes_G* includes the students who answered "yes" to the respective variable, and *No_G* comprises the students who responded "no".

The results of MGA are summarised in Table 9. Our analysis shows that *prior entrepreneurship experience* and *entrepreneur in immediate family* had no significant moderating effects on our model. We found that those students who had a businessperson in their immediate family had a significantly higher path coefficient for the association between entrepreneurial inspiration and EA compared to the students who did not. This means that the *businessperson in immediate family* variable could significantly moderate the link between entrepreneurial inspiration and students' EAs (*p-value=0.024*). It is worth noting that the *businessperson in immediate family* variable had no significant moderating effect on other path coefficients.

We also performed the Mann-Whitney U test (Howell, 2012) to examine the impacts of the demographic variables on the EAs of students and their EIs measured in two times: at the start of the OzGirlsEntrepreneurship workshops (*time1*) and the end of the OzGirlsEntrepreneurship workshops (*time2*). Table 9 shows that the students who had a businessperson in their immediate family exhibited significantly higher intention in both time1 (*U=3813.0, N1=103, N2=90, p-value=0.026, r=0.152*) and time2 (*U=3834.0, N1=103, N2=90, p-value=0.028, r=0.148*) (B. H. Cohen, 2008) compared to the students who did not have a businessperson in their immediate family.

Table 9. Moderating effects of demographics

| Moderating Factors | H1: Attitude → Intention | H3: Inspiration → Attitude | H4: Inspiration → Intention | H6: Learning → Attitude | H7: Learning → Intention | Intention (*time1*) | Intention (*time2*) | Attitude |
|---|---|---|---|---|---|---|---|---|
| Prior entrepreneurship experience | No Diff. | No Diff. | No Diff. | No Diff. | No Diff. | No Diff. | No Diff. | No Diff. |
| Businessperson in immediate family | No Diff. | **Yes_G > No_G*** | No Diff. | No Diff. | No Diff. | **Yes_G > No_G*** | **Yes_G> No_G*** | No Diff. |
| Entrepreneur in immediate family | No Diff. | No Diff. | No Diff. | No Diff. | No Diff. | No Diff. | No Diff. | No Diff. |

* Significant at p-value < 0.05

## 5 Discussion and Conclusions

This study has analysed the effect of a one-day female-focused STEM-based entrepreneurship program (i.e., the OzGirlsEntrepreneurship program) on the entrepreneurial intention of secondary school female students through a theoretical model. We used two surveys to collect data from 193 Year 10 female student respondents, who participated in the OzGirlsEntrepreneurship program. The participants aged 14-16 old and came from 44 schools (Government, Independent, Catholic schools) in Victoria state, Australia. We believe our findings contribute to the body of the research on entrepreneurship, on entrepreneurship education more specifically, for (female) teenagers, who have been less researched compared to adults. In the rest of this section, we first discuss our key findings in the context of existing literature. Then, we provide some implications for researchers, policymakers, and practitioners, followed by some suggestions and recommendations to design and organise entrepreneurship education programs for teenagers. Finally, we report the limitations of our study and potential research directions.

### 5.1 Discussion on the Findings

**H1** (Attitude → Intention) was accepted, suggesting a positive substantial link between entrepreneurial attitude and entrepreneurial intent. The effect size is 0.665, which implies a large effect. In this regard, our result supports the findings of a few studies on secondary school students (Do Paco, Ferreira, Raposo, Rodrigues, & Dinis, 2011; Purwana, Suhud, & Rahayu, 2017) and matches the results of many studies that targeted adults, for instance (Karimi et al., 2016; Nguyen et al., 2019). In this respect, after the OzGirlsEntrepreneurship program, 68.3% of students perceived entrepreneurship as an attractive career and reported that they see the prospect of establishing their own business in the future. However, the OzGirlsEntrepreneurship program participants, on average, demonstrated a slightly higher entrepreneurial attitude (3.927 out of 5) than entrepreneurial intention (3.772 out of 5).

The OzGirlsEntrepreneurship program was deemed helpful to significantly enhance the intent of female students to become an entrepreneur, rising from the average value of 3.37 out of 5 to 3.77. Hence, **H2** (Education → Intention) was accepted. The size effect is 0.413, which indicates its effect is large. Our finding supports (Ni & Ye, 2018; José C Sánchez, 2013), which revealed a substantial positive influence of entrepreneurship education on the entrepreneurial intent of secondary school students. Rodrigues et al. (2012) observed that their entrepreneurship programs did not statistically change teenagers' intentions towards pursuing an entrepreneurial career. Perhaps surprisingly, Oosterbeek et al. (2010) and Huber, Sloof, and Van Praag (2014) observed this effect was negative. Given that some studies, such as (Barba-Sánchez & Atienza-Sahuquillo, 2018; Karimi et al., 2016) in the university context and (Huber et al., 2014; Oosterbeek et al., 2010) at the lower level of education systems, could not achieve the intended goals from their respective entrepreneurship education programs, we speculate that these diverse findings may stem from the pedagogical approaches and content of entrepreneurship programs and different cultural contexts. Future research needs to target various cultural contexts to improve generalisability and analyse the characteristics of different types of entrepreneurship programs in shaping and promoting the entrepreneurial attitudes and intentions of secondary school students (Moberg, 2014).

The entrepreneurship research in secondary education does not, however, investigate the benefits (i.e., inspiration and learning) offered by an entrepreneurship program. Hence, our study is the first one, which investigated entrepreneurial inspiration and learning in secondary education. We observed that there is a statistically positive relation between entrepreneurial inspiration and entrepreneurial attitude, which led to accepting **H3** (Inspiration → Attitude) (i.e., the effect size is small). However, we witnessed no evidence that entrepreneurial inspiration is significantly correlated with entrepreneurial intent and **H4** (Inspiration → Intention) was rejected. However, this correlation could be significant when entrepreneurial attitude plays a mediating role, leading to supporting **H5** (Inspiration → Attitude → Intention).

The hypotheses **H6** (Learning → Attitude), **H7** (Learning → Intention), and **H8** (Learning → Attitude → Intention) assess to what extent the perceptions of students on their learning from the OzGirlsEntrepreneurship program are correlated with their entrepreneurial attitude and intention. Our study found that the students who had greater learning from the OzGirlsEntrepreneurship program indicated a higher attitude towards entrepreneurship (**H6** was accepted). We had no evidence to suggest that students' perceived learning from the OzGirlsEntrepreneurship program is statistically associated with their intentions to be self-employment (**H7** was rejected). Similar to **H5**, we found that entrepreneurial attitude can have a mediation role in linking entrepreneurial learning and intention (**H8** was accepted).

Finally, we analysed the effect of several demographics on our theoretical model. Generally speaking, we observed that demographics had no significant impact on our model. We only found that having a businessperson in the immediate family statistically increased the strength of the link between entrepreneurial inspiration and entrepreneurial attitude.

### 5.2 Implications

Our research findings draw out implications that can support the future design and delivery of entrepreneurship education programs across a wide span of educational sectors. In particular, a key factor in the development of entrepreneurial intent in young female students is associated with building an entrepreneurial attitude or 'mindset' and considering how this aligns to soft-skills development in the areas of creative problem-solving that are perceived to broadly benefit career and academic outcomes. Other factors include positive role modelling experiences and the strengthening of peer connections, and an alignment to meaningful and real-world initiatives.

The primary allure for young women in fostering and developing an entrepreneurial attitude lies in the learners seeing potential for growth in the areas of creative thinking, risk-taking, problem-solving and leadership development. Of particular significance to both fields are the positive findings associated with the development of entrepreneurial "attitudes". The foregrounding of the attitudes is somewhat of a recent trend in educational contexts, especially when compared to traditional educational foci on content, knowledge, and skills development (Huber et al., 2014). The results of this study that indicate higher levels of entrepreneurial attitude in comparison to entrepreneurial intent showcase the importance for educational designers, teachers, and mentors to consider improved practices in designing for, teaching, integrating, and assessing attitudinal aspects throughout future entrepreneurship education programs. This includes focusing on the development of student attitudes toward creative risk-taking, motivation and leadership, the problem-solving processes, and failure (Jose C Sánchez & Licciardello, 2012). Such sentiments are echoed in works on entrepreneurship education, particularly in European contexts (Huber et al., 2014; Testa & Frascheri, 2015). The development of entrepreneurial attitudes, or "mindsets", is especially important when considering the factors affecting female students' intent toward becoming an entrepreneur.

Whilst the OzGirlsEntrepreneurship research findings did not reveal strong correlations between having a business person in the immediate family and future entrepreneurial intent, they draw out a thread that highlights the importance of meaningful human connections to motivate and persuade intent, as well as emphasise the emotional significance of such programs (Karimi et al., 2016; Nabi et al., 2018). In this study, this correlation was found to have relevance in the development of entrepreneurial attitude when aligned to a positive mentoring, modelling, or a mediating role. The strong correlation between these two factors highlights a shift away from traditional skills and knowledge-based learning styles to those that foster the development of attitudes, emotions, and mindsets more significantly (do Paço, Ferreira, Raposo, Rodrigues, & Dinis, 2011).

Further to this, it is important for young females that their entrepreneurial intent is aligned to relevant, societal problems and a broader global purpose and overall positive impact on the world. Despite the OzGirlsEntrepreneurship program focusing on a female, secondary school-aged cohort in a technology-focused context, a presumption can be made that this emphasis on real-world problem-solving would be applicable to a range of cohorts and demographics. Importantly, the weight of the results highlights the significance of these findings in the future development and delivery of entrepreneurship education programs for groups in upper-primary school, pre-university, as well as at undergraduate levels of study. Such findings have implications across future research and practice in the entrepreneurship education sector, but also more broadly in relation to STEM-based learning.

With these factors in mind, it is important to acknowledge the relevance of pedagogical strategies to support the design and delivery of entrepreneurship educational experiences that promote attitude and mindset development. This includes but is not limited to pedagogical strategies such as problem-based learning, technology-based simulations, experiential learning, and mentoring (Pittaway & Cope, 2007; Stewart & Knowles, 2003). Such considerations are of particular relevance when considered in relation to the importance of attitudinal development in students and how such attitudes might be designed for, delivered, and, more importantly, assessed within academic or educational frameworks.

**5.3 Limitations and Future Research**

The design and results of this study may have some limitations. Measuring entrepreneurial attitude, intention, and learning with single-item indicators has limitations. Single-item indicators are often used to measure these constructs validly, e.g., (Krueger et al., 2000; Wilson et al., 2007), and researchers in other fields have shown that single-item indicators can produce the comparable level of validity and reliability to multiple-item indicators (Bergkvist & Rossiter, 2007; Bowling, 2005; Petrescu, 2013). Despite this, utilising multiple-item measures to assess these constructs in future research will assist with minimising measurement error (Krueger et al., 2000). Our findings may not be generalised to students of any gender or in other cultural contexts (e.g., male students in other regions), as our sample was made of 193 female students who came from 44 schools in the state of Victoria in Australia. The generalizability of our findings can also be limited by the fact that the majority of students were academically high-performers (See Fig. 2). This means that the OzGirlsEntrepreneurship program with a broader spectrum of students could produce different results. Another threat to our findings comes from the recruitment method.

Participating in the OzGirlsEntrepreneurship program was not compulsory for students, indicating the participating students were more likely interested in the entrepreneurial activities. Therefore, self-selection bias may have occurred (Lavrakas, 2008). Future research should focus on teenagers who are obliged to attend an entrepreneurship education program regardless of interest. Considering the age of students, they might have responded to some of the survey questions (e.g., "*The OzGirlsEntrepreneurship program has increased my understanding of the actions that someone has to take in order to start an entrepreneurial career*") in a way that are more acceptable for us as researchers (i.e., social desirability bias) (Furnham, 1986). We took the following mitigation strategies for this threat: first, students were ensured that both surveys are anonymous, and they would not be identifiable in the potential publications in the future. Second, at the beginning of each OzGirlsEntrepreneurship workshop, we told students that they would not be judged based on their responses to the surveys.

Given the significant role of pedagogical strategies employed in entrepreneurship programs in cultivating attitude and mindset, an area of ongoing interest and future research might, therefore, focus on how the development of entrepreneurship programs that allow such attitudes to lead not only to an increase in entrepreneurial intent but also to positive improvement in academic achievement (Huber et al., 2014; Johansen, 2014). Such studies would, however, require further inquiry on what additional socio-demographic and educational variables that might interact with entrepreneurial attitude. This might include considering the relationship between subject-specific performance or preference (e.g., in STEM) and entrepreneurial attitude or entrepreneurial intention, alongside a more granular inquiry into the background aspects of the student (e.g., ethnicity and class), to consider how these factors play a role. To reiterate, there is a need to embrace various cultural contexts to explore the bespoke program development of entrepreneurship education aimed at shaping the entrepreneurial attitudes and intentions of secondary school students (Moberg, 2014). Further, given the demographic, high academic achievement levels, and STEM-focus of the OzGirlsEntrepreneurship student cohort, additional studies are also recommended to better understand the link between entrepreneurship and STEM education to academic achievement and the pursuit of entrepreneurial ventures.

There is limited longitudinal research investigating the impacts of entrepreneurship education programs on students in general and on secondary school students in particular (Elert et al., 2015). As a result, we plan to analyse the longitudinal influence of the OzGirlsEntrepreneurship program on the participating students' entrepreneurial attitude and intention, self-employment, and undertaking a business venture.

# Acknowledgments

This work is funded by the Australian Government, Department of Industry, Innovation and Science, Grant No. WISE64905. The authors would like to thank all participants in this study.

# Appendix. Survey Questions

Table 10. Pre-workshop Survey Questions

| Questions | Scale |
| --- | --- |
| **Demographic Questions** | |
| "Is there any person in your immediate family who was/is an entrepreneur?" | Yes / No / I do not know |
| "If you answered **Yes** to the previous question, how many (roughly) people in your family were/are an entrepreneur?" | Free text |
| "Is there any person in your immediate family who has owned-managed/is owning-managing a company or a business?" | Yes / No / I do not know |
| "If you answered **Yes** to the previous question, how many (roughly) people in your family have owned-managed/are owning-managing a company or a business?" | Free text |
| "Have you engaged in any prior entrepreneurship-related training?" | Yes / No |
| "If you answered "Yes" to the previous question, how long (roughly) have you engaged in the entrepreneurship-related training?" | Free text |
| "In what mark range is your overall academic performance in class?" | "90%-100%" / "80%-89.9%" / "70%-79.9%" / "60%-69.9%" / "50%-59.9%" / "0%-49.9%" |
| **Entrepreneurship Question** | |
| "I would like to start my own business one day." | "Strongly agree" / "Agree" / "Neutral" / "Disagree" / "Strongly disagree" |

Table 11. Post-workshop Survey Questions

| Questions | Scale |
| --- | --- |
| **Entrepreneurship Questions** | |
| "Participating in the OzGirlsEntrepreneurship program has increased my understanding of the actions that someone has to take in order to start an entrepreneurial career." | "Strongly agree" / "Agree" / "Neutral" / "Disagree" / "Strongly disagree" |
| "Having participated in OzGirlsEntrepreneurship program, I would like to start my own business one day." | "Strongly agree" / "Agree" / "Neutral" / "Disagree" / "Strongly disagree" |
| "Having participated in OzGirlsEntrepreneurship program, a career as an entrepreneur is attractive to me." | "Strongly agree" / "Agree" / "Neutral" / "Disagree" / "Strongly disagree" |
| *"How did the following experiences from the OzGirlsEntrepreneurship program impact your mind towards taking a career in entrepreneurship?"* | |
| "Seeing role models working in entrepreneurship." | "Very negative" / "Negative" / "Neutral" / "Positive" / "Very positive" |
| "Working with my mentor." | "Very negative" / "Negative" / "Neutral" / "Positive" / "Very positive" |
| "Working with my peers." | "Very negative" / "Negative" / "Neutral" / "Positive" / "Very positive" |
| "Learning new ways of thinking." | "Very negative" / "Negative" / "Neutral" / "Positive" / "Very positive" |
| "Practising market research." | "Very negative" / "Negative" / "Neutral" / "Positive" / "Very positive" |
| "Delivering a pitch." | "Very negative" / "Negative" / "Neutral" / "Positive" / "Very positive" |
| "Connecting technology to a real-world idea." | "Very negative" / "Negative" / "Neutral" / "Positive" / "Very positive" |
| "Hands-on coding." | "Very negative" / "Negative" / "Neutral" / "Positive" / "Very positive" |
| "Getting feedback on my ideas." | "Very negative" / "Negative" / "Neutral" / "Positive" / "Very positive" |

# References

Adekiya, A. A., & Ibrahim, F. (2016). Entrepreneurship intention among students. The antecedent role of culture and entrepreneurship training and development. *The International Journal of Management Education, 14*(2), 116-132. doi:https://doi.org/10.1016/j.ijme.2016.03.001

Ahl, H. (2006). Why research on women entrepreneurs needs new directions. *Entrepreneurship Theory and Practice, 30*(5), 595-621.

Ahmad, F., Jhajj, A. K., Stewart, D. E., Burghardt, M., & Bierman, A. S. (2014). Single item measures of self-rated mental health: a scoping review. *BMC health services research, 14*(1), 1-11.

Ajzen, I. (1991). The theory of planned behavior. *Organizational Behavior and Human Decision Processes, 50*(2), 179-211. doi:https://doi.org/10.1016/0749-5978(91)90020-T

Ali, I., Ali, M., Badghish, S., & Baazeem, T. (2018). Examining the Role of Childhood Experiences in Developing Altruistic and Knowledge Sharing Behaviors among Children in Their Later Life: A Partial Least Squares (PLS) Path Modeling Approach. *Sustainability, 10*(2), 292.

Allen, E., Langowitz, N., & Minniti, M. (2007). The 2006 global entrepreneurship monitor special topic report: women in entrepreneurship. *Center for Women Leadership, Babson College*.

Armuña, C., Ramos, S., Juan, J., Feijóo, C., & Arenal, A. (2020). From stand-up to start-up: exploring entrepreneurship competences and STEM women's intention. *International Entrepreneurship and Management Journal*, 1-24.

Arranz, N., Ubierna, F., Arroyabe, M. F., Perez, C., & Fdez. de Arroyabe, J. C. (2017). The effect of curricular and extracurricular activities on university students' entrepreneurial intention and competences. *Studies in Higher Education, 42*(11), 1979-2008. doi:10.1080/03075079.2015.1130030

Ashcraft, C., & Blithe, S. (2010). Women in IT: The facts. *National Center for Women & Information Technology*.

Barba-Sánchez, V., & Atienza-Sahuquillo, C. (2018). Entrepreneurial intention among engineering students: The role of entrepreneurship education. *European Research on Management and Business Economics, 24*(1), 53-61. doi:https://doi.org/10.1016/j.iedeen.2017.04.001

Barclay, C., Donalds, C., & Osei-Bryson, K.-M. (2018). Investigating critical success factors in online learning environments in higher education systems in the Caribbean. *Information Technology for Development, 24*(3), 582-611.

Barker, L. J., & Aspray, W. (2006). *The state of research on girls and IT*: na.

Baron, R. M., & Kenny, D. A. (1986). The moderator–mediator variable distinction in social psychological research: Conceptual, strategic, and statistical considerations. *Journal of personality and social psychology, 51*(6), 1173.

Beliaeva, T., Laskovaia, A., & Shirokova, G. (2017). Entrepreneurial learning and entrepreneurial intentions: A cross-cultural study of university students. *European Journal of International Management, 11*(5), 606-632.

Bergkvist, L., & Rossiter, J. R. (2007). The predictive validity of multiple-item versus single-item measures of the same constructs. *Journal of marketing research, 44*(2), 175-184.

Bird, B. (1988). Implementing entrepreneurial ideas: The case for intention. *Academy of management Review, 13*(3), 442-453.

Boddington, M., & Barakat, S. (2018). Exploring alternative gendered social structures within entrepreneurship education: notes from a women's-only enterprise programme in the United Kingdom. In *Women Entrepreneurs and the Myth of 'Underperformance'*: Edward Elgar Publishing.

Bosma, N., Hessels, J., Schutjens, V., Praag, M. V., & Verheul, I. (2012). Entrepreneurship and role models. *Journal of Economic Psychology, 33*(2), 410-424. doi:https://doi.org/10.1016/j.joep.2011.03.004

Bosma, N., & Kelley, D. (2019). *Global Entrepreneurship Monitor 2018/2019 Global Report*. Retrieved from https://www.gemconsortium.org/file/open?fileId=50213:

Bosma, N., & Levie, J. (2010). Global entrepreneurship monitor: 2009 global report.

Bowling, A. (2005). Just one question: If one question works, why ask several? In: BMJ Publishing Group Ltd.

Boyd, N. G., & Vozikis, G. S. (1994). The influence of self-efficacy on the development of entrepreneurial intentions and actions. *Entrepreneurship Theory and Practice, 18*(4), 63-77.

Cárcamo-Solís, M. d. L., Arroyo-López, M. d. P., Alvarez-Castañón, L. d. C., & García-López, E. (2017). Developing entrepreneurship in primary schools. The Mexican experience of "My first enterprise: Entrepreneurship by playing". *Teaching and Teacher Education, 64*, 291-304. doi:https://doi.org/10.1016/j.tate.2017.02.013

Catherine Ashcraft, E. E., and Michelle Friend. (2012). Girls in IT: The facts. Boulder, CO: National Centre for Women & Information Technology (NCWIT). In.

Chin, W. W. (2010). How to write up and report PLS analyses. In *Handbook of partial least squares* (pp. 655-690): Springer.

Cochran, S. L. (2019). What's Gender Got to Do with It? The Experiences of US Women Entrepreneurship Students. *Journal of Small Business Management, 57*(sup1), 111-129.

Cohen, B. H. (2008). *Explaining psychological statistics*: John Wiley & Sons.

Cohen, J. (1988). Statistical power analysis for the behavioral sciences. Abingdon. In: United Kingdom: Routledge.

Cui, J., Sun, J., & Bell, R. (2019). The impact of entrepreneurship education on the entrepreneurial mindset of college students in China: The mediating role of inspiration and the role of educational attributes. *The International Journal of Management Education*, 100296. doi:https://doi.org/10.1016/j.ijme.2019.04.001

Dabic, M., Daim, T., Bayraktaroglu, E., Novak, I., & Basic, M. (2012). Exploring gender differences in attitudes of university students towards entrepreneurship. *International Journal of Gender and Entrepreneurship*.

Daim, T., Dabic, M., & Bayraktaroglu, E. (2016). Students' entrepreneurial behavior: international and gender differences. *Journal of Innovation and Entrepreneurship, 5*(1), 19.

Degeorge, J. M., & Fayolle, A. (2008). Is entrepreneurial intention stable through time? First insights from a sample of French students. *International Journal of Entrepreneurship and Small Business, 5*(1), 7-27.

Deutskens, E., De Ruyter, K., Wetzels, M., & Oosterveld, P. (2004). Response rate and response quality of internet-based surveys: An experimental study. *Marketing letters, 15*(1), 21-36.

Devine, J., Finney, J., de Halleux, P., Moskal, M., Ball, T., & Hodges, S. (2019). MakeCode and CODAL: intuitive and efficient embedded systems programming for education. *Journal of Systems Architecture, 98*, 468-483.

Díaz-García, M. C., & Jiménez-Moreno, J. (2010). Entrepreneurial intention: the role of gender. *International Entrepreneurship and Management Journal, 6*(3), 261-283.

Do Paço, A., Ferreira, J., Raposo, M., Rodrigues, R. G., & Dinis, A. (2011). Entrepreneurial intention among secondary students: findings from Portugal. *International Journal of Entrepreneurship and Small Business, 13*(1), 92-106.

do Paço, A. M. F., Ferreira, J. M., Raposo, M., Rodrigues, R. G., & Dinis, A. (2011). Behaviours and entrepreneurial intention: Empirical findings about secondary students. *Journal of International Entrepreneurship, 9*(1), 20-38.

Dyer, W. G. (1995). Toward a Theory of Entrepreneurial Careers. *Entrepreneurship Theory and Practice, 19*(2), 7-21. doi:10.1177/104225879501900202

Elert, N., Andersson, F. W., & Wennberg, K. (2015). The impact of entrepreneurship education in high school on long-term entrepreneurial performance. *Journal of Economic Behavior & Organization, 111*, 209-223. doi:https://doi.org/10.1016/j.jebo.2014.12.020

Elliott, C., Mavriplis, C., & Anis, H. (2020). An entrepreneurship education and peer mentoring program for women in STEM: mentors' experiences and perceptions of entrepreneurial self-efficacy and intent. *International Entrepreneurship and Management Journal*. doi:10.1007/s11365-019-00624-2

Engle, R. L., Dimitriadi, N., Gavidia, J. V., Schlaegel, C., Delanoe, S., Alvarado, I., . . . Wolff, B. (2010). Entrepreneurial intent: A twelve-country evaluation of Ajzen's model of planned behavior. *International Journal of Entrepreneurial Behavior & Research, 16*(1), 35-57.

Erikson, T. (2003). Towards a taxonomy of entrepreneurial learning experiences among potential entrepreneurs. *Journal of Small Business and Enterprise Development, 10*(1), 106-112.

Fagerberg, J., & Srholec, M. (2009). *Knowledge, Capabilities, and the Poverty Trap: The Complex Interplay Between Technological, Social, and Geographical Factors* (Working Paper No.24/2009). Retrieved from Centre for Technology, Innovation and Culture, University of Oslo:

Farnell, T., Heder, E., & Ljubić, M. (2015). Entrepreneurship Education in the European Union: An Overview Of Policies And Practice. In: Zagreb: South East European Centre for Entrepreneurial Learning.

Fayolle, A., & Gailly, B. (2015). The impact of entrepreneurship education on entrepreneurial attitudes and intention: Hysteresis and persistence. *Journal of Small Business Management, 53*(1), 75-93.

Fayolle, A., & Liñán, F. (2014). The future of research on entrepreneurial intentions. *Journal of Business Research, 67*(5), 663-666. doi:https://doi.org/10.1016/j.jbusres.2013.11.024

Fellnhofer, K. (2019). Toward a taxonomy of entrepreneurship education research literature: A bibliometric mapping and visualization. *Educational Research Review*.

Ferri, L., Ginesti, G., Spanò, R., & Zampella, A. (2018). Exploring the Entrepreneurial Intention of Female Students in Italy. *Journal of Open Innovation: Technology, Market, and Complexity, 4*(3), 27.

Filion, L. J. (1994). Ten steps to entrepreneurial teaching. *Journal of Small Business & Entrepreneurship, 11*(3), 68-78.

Fornell, C., & Larcker, D. F. (1981). Evaluating structural equation models with unobservable variables and measurement error. *Journal of marketing research, 18*(1), 39-50.

Furnham, A. (1986). Response bias, social desirability and dissimulation. *Personality and Individual Differences, 7*(3), 385-400. doi:https://doi.org/10.1016/0191-8869(86)90014-0

Gasse, Y. (1985). A strategy for the promotion and identification of potential entrepreneurs at the secondary school level. *Frontiers of entrepreneurship*, 538-559.

Gefen, D., & Straub, D. (2005). A practical guide to factorial validity using PLS-Graph: Tutorial and annotated example. *Communications of the Association for Information systems, 16*(1), 5.

Geisser, S. (1974). A predictive approach to the random effect model. *Biometrika, 61*(1), 101-107.

Hair, J., Ringle, C., & Sarstedt, M. (2011). PLS-SEM: Indeed a silver bullet. *Journal of Marketing theory and Practice, 19*(2), 139-152.

Hair, J., Risher, J., Sarstedt, M., & Ringle, C. M. (2019). When to use and how to report the results of PLS-SEM. *European Business Review, 31*(1), 2-24.
Hair, J. F., Ringle, C. M., & Sarstedt, M. (2012). Partial least squares: the better approach to structural equation modeling? *Long Range Planning, 45*(5-6), 312-319.
Hair Jr, J. F., Hult, G. T. M., Ringle, C., & Sarstedt, M. (2016). *A primer on partial least squares structural equation modeling (PLS-SEM)*: Sage publications.
Halabisky, D. (2018). *Policy Brief on Women's Entrepreneurship* (OECD SME and Entrepreneurship Papers No. 8). Retrieved from https://www.oecd.org/cfe/smes/Policy-Brief-on-Women-s-Entrepreneurship.pdf:
Henseler, J., & Chin, W. W. (2010). A comparison of approaches for the analysis of interaction effects between latent variables using partial least squares path modeling. *Structural Equation Modeling, 17*(1), 82-109.
Henseler, J., & Fassott, G. (2010). Testing moderating effects in PLS path models: An illustration of available procedures. In *Handbook of partial least squares* (pp. 713-735): Springer.
Henseler, J., Ringle, C. M., & Sarstedt, M. (2015). A new criterion for assessing discriminant validity in variance-based structural equation modeling. *Journal of the Academy of Marketing Science, 43*(1), 115-135. doi:10.1007/s11747-014-0403-8
Hmelo-Silver, C. E. (2004). Problem-based learning: What and how do students learn? *Educational Psychology Review, 16*(3), 235-266.
Howell, D. C. (2012). *Statistical methods for psychology*: Cengage Learning.
Huber, L. R., Sloof, R., & Van Praag, M. (2014). The effect of early entrepreneurship education: Evidence from a field experiment. *European Economic Review, 72*, 76-97. doi:https://doi.org/10.1016/j.euroecorev.2014.09.002
Johannisson, B. (1991). University training for entrepreneurship: Swedish approaches. *Entrepreneurship & Regional Development, 3*(1), 67-82. doi:10.1080/08985629100000005
Johansen, V. (2014). Entrepreneurship education and academic performance. *Scandinavian Journal of Educational Research, 58*(3), 300-314.
Jöreskog, K. G. (1970). A general method for estimating a linear structural equation system. *ETS Research Bulletin Series, 1970*(2), i-41.
Karimi, S., Biemans, H. J. A., Lans, T., Chizari, M., & Mulder, M. (2016). The Impact of Entrepreneurship Education: A Study of Iranian Students' Entrepreneurial Intentions and Opportunity Identification. *Journal of Small Business Management, 54*(1), 187-209. doi:10.1111/jsbm.12137
Kariv, D. (2013). *Female entrepreneurship and the new venture creation: An international overview*: Routledge.
Katz, J., & Gartner, W. B. (1988). Properties of emerging organizations. *Academy of management Review, 13*(3), 429-441.
Kolvereid, L., & Isaksen, E. (2006). New business start-up and subsequent entry into self-employment. *Journal of Business Venturing, 21*(6), 866-885.
KritiKoS, A. S. (2014). Entrepreneurs and their impact on jobs and economic growth. *IZA World of Labor*.
Krueger, N. F., Reilly, M. D., & Carsrud, A. L. (2000). Competing models of entrepreneurial intentions. *Journal of Business Venturing, 15*(5), 411-432. doi:https://doi.org/10.1016/S0883-9026(98)00033-0
Kuschel, K., Ettl, K., Díaz-García, C., & Alsos, G. A. (2020). Stemming the gender gap in STEM entrepreneurship–insights into women's entrepreneurship in science, technology, engineering and mathematics. *International Entrepreneurship and Management Journal*, 1-15.
Lans, T., Blok, V., & Gulikers, J. (2015). Show me your network and I'll tell you who you are: social competence and social capital of early-stage entrepreneurs. *Entrepreneurship & Regional Development, 27*(7-8), 458-473. doi:10.1080/08985626.2015.1070537
Lavrakas, P. J. (2008). *Encyclopedia of survey research methods*: Sage Publications.
Liñán, F., & Chen, Y. W. (2009). Development and cross–cultural application of a specific instrument to measure entrepreneurial intentions. *Entrepreneurship Theory and Practice, 33*(3), 593-617.
Liñán, F., & Fayolle, A. (2015). A systematic literature review on entrepreneurial intentions: citation, thematic analyses, and research agenda. *International Entrepreneurship and Management Journal, 11*(4), 907-933. doi:10.1007/s11365-015-0356-5
Lortie, J., & Castogiovanni, G. (2015). The theory of planned behavior in entrepreneurship research: what we know and future directions. *International Entrepreneurship and Management Journal, 11*(4), 935-957.
Low, K., Yoon, M., Roberts, B. W., & Rounds, J. (2005). The stability of vocational interests from early adolescence to middle adulthood: a quantitative review of longitudinal studies. *Psychological bulletin, 131*(5), 713.
Maresch, D., Harms, R., Kailer, N., & Wimmer-Wurm, B. (2016). The impact of entrepreneurship education on the entrepreneurial intention of students in science and engineering versus business studies university programs. *Technological forecasting and social change, 104*, 172-179.
Margolis, J., Goode, J., & Bernier, D. (2011). The need for computer science. *Educational Leadership, 68*(5), 68-72.
Marlow, S., & Swail, J. (2014). Gender, risk and finance: why can't a woman be more like a man? *Entrepreneurship & Regional Development, 26*(1-2), 80-96.
Martin, B. C., McNally, J. J., & Kay, M. J. (2013). Examining the formation of human capital in entrepreneurship: A meta-analysis of entrepreneurship education outcomes. *Journal of Business Venturing, 28*(2), 211-224. doi:https://doi.org/10.1016/j.jbusvent.2012.03.002
Mascarenhas, D. D., & Vander Veer, S. (2014). Women, innovation, and literature. *Journal of Innovation and Entrepreneurship, 3*(1), 7.
McDonald, J. H. (2009). *Handbook of biological statistics* (Vol. 2): sparky house publishing Baltimore, MD.
Meade, A. W., & Craig, S. B. (2012). Identifying careless responses in survey data. *Psychological methods, 17*(3), 437.
Menzies, T. V., & Tatroff, H. (2006). The Propensity of Male vs. Female Students To Take Courses and Degree Concentrations in Entrepreneurship1. *Journal of Small Business & Entrepreneurship, 19*(2), 203-223.
Miliszewska, I., & Sztendur, E. M. (2010). Interest in ICT studies and careers: Perspectives of secondary school female students from low socioeconomic backgrounds. *Interdisciplinary Journal of Information, Knowledge, and Management, 5*, 237-260.
Moberg, K. (2014). Two approaches to entrepreneurship education: The different effects of education for and through entrepreneurship at the lower secondary level. *The International Journal of Management Education, 12*(3), 512-528.
Moriano, J. A., Gorgievski, M., Laguna, M., Stephan, U., & Zarafshani, K. (2012). A cross-cultural approach to understanding entrepreneurial intention. *Journal of career development, 39*(2), 162-185.
Nabi, G., Holden, R., & Walmsley, A. (2010). Entrepreneurial intentions among students: towards a re-focused research agenda. *Journal of Small Business and Enterprise Development*.
Nabi, G., Walmsley, A., Liñán, F., Akhtar, I., & Neame, C. (2018). Does entrepreneurship education in the first year of higher education develop entrepreneurial intentions? The role of learning and inspiration. *Studies in Higher Education, 43*(3), 452-467. doi:10.1080/03075079.2016.1177716
Neumeyer, X., & Santos, S. C. (2020). A lot of different flowers make a bouquet: The effect of gender composition on technology-based entrepreneurial student teams. *International Entrepreneurship and Management Journal, 16*(1), 93-114. doi:10.1007/s11365-019-00603-7
Nguyen, A. T., Do, T. H. H., Vu, T. B. T., Dang, K. A., & Nguyen, H. L. (2019). Factors affecting entrepreneurial intentions among youths in Vietnam. *Children and Youth Services Review, 99*, 186-193. doi:https://doi.org/10.1016/j.childyouth.2019.01.039
Ni, H., & Ye, Y. (2018). Entrepreneurship Education Matters: Exploring Secondary Vocational School Students' Entrepreneurial Intention in China. *The Asia-Pacific Education Researcher, 27*(5), 409-418.
Nitzl, C., Roldan, J. L., & Cepeda, G. (2016). Mediation analysis in partial least squares path modeling: Helping researchers discuss more sophisticated models. *Industrial management & data systems, 116*(9), 1849-1864.
Nowiński, W., & Haddoud, M. Y. (2019). The role of inspiring role models in enhancing entrepreneurial intention. *Journal of Business Research, 96*, 183-193. doi:https://doi.org/10.1016/j.jbusres.2018.11.005
Nowiński, W., Haddoud, M. Y., Lančarič, D., Egerová, D., & Czeglédi, C. (2019). The impact of entrepreneurship education, entrepreneurial self-efficacy and gender on entrepreneurial intentions of university students in the Visegrad countries. *Studies in Higher Education, 44*(2), 361-379. doi:10.1080/03075079.2017.1365359
O'Connor, A. (2013). A conceptual framework for entrepreneurship education policy: Meeting government and economic purposes. *Journal of Business Venturing, 28*(4), 546-563.
Oosterbeek, H., van Praag, M., & Ijsselstein, A. (2010). The impact of entrepreneurship education on entrepreneurship skills and motivation. *European Economic Review, 54*(3), 442-454. doi:https://doi.org/10.1016/j.euroecorev.2009.08.002
Ozaralli, N., & Rivenburgh, N. K. (2016). Entrepreneurial intention: antecedents to entrepreneurial behavior in the USA and Turkey. *Journal of Global Entrepreneurship Research, 6*(1), 3.
Papastergiou, M. (2008). Are computer science and information technology still masculine fields? High school students' perceptions and career choices. *Computers & Education, 51*(2), 594-608.

Peterman, N. E., & Kennedy, J. (2003). Enterprise education: Influencing students' perceptions of entrepreneurship. *Entrepreneurship Theory and Practice, 28*(2), 129-144.

Petrescu, M. (2013). Marketing research using single-item indicators in structural equation models. *Journal of marketing analytics, 1*(2), 99-117.

Pittaway, L., & Cope, J. (2007). Entrepreneurship education: A systematic review of the evidence. *International small business journal, 25*(5), 479-510.

Poggesi, S., Mari, M., De Vita, L., & Foss, L. (2020). Women entrepreneurship in STEM fields: literature review and future research avenues. *International Entrepreneurship and Management Journal, 16*.1. doi:10.1007/s11365-019-00599-0

Purwana, D., Suhud, U., & Rahayu, S. (2017). Entrepreneurial intention of secondary and tertiary students: Are they different. *International Journal of Economic Research, 14*(18), 69-81.

Rodrigues, R. G., Dinis, A., do Paço, A., Ferreira, J., & Raposo, M. (2012). The effect of an entrepreneurial training programme on entrepreneurial traits and intention of secondary students. *Entrepreneurship–Born, made and educated*, 77-92.

Sánchez, J. C. (2013). The impact of an entrepreneurship education program on entrepreneurial competencies and intention. *Journal of Small Business Management, 51*(3), 447-465.

Sánchez, J. C., & Licciardello, O. (2012). Gender differences and attitudes in entrepreneurial intentions: the role of career choice. *Journal of Women's Entrepreneurship and Education*(1-2), 7-27.

Sarstedt, M., Henseler, J., & Ringle, C. M. (2011). Multigroup analysis in partial least squares (PLS) path modeling: Alternative methods and empirical results. In *Measurement and research methods in international marketing* (pp. 195-218): Emerald Group Publishing Limited.

Shapero, A., & Sokol, L. (1982). The social dimensions of entrepreneurship. *Encyclopedia of entrepreneurship*, 72-90.

Sharma, G. G., & Stol, K.-J. (2020). Exploring onboarding success, organizational fit, and turnover intention of software professionals. *Journal of Systems and Software, 159*, 110442. doi:https://doi.org/10.1016/j.jss.2019.110442

Shinnar, R. S., Giacomin, O., & Janssen, F. (2012). Entrepreneurial perceptions and intentions: The role of gender and culture. *Entrepreneurship Theory and Practice, 36*(3), 465-493.

Simard, C. (2008). *Climbing the technical ladder: Obstacles and solutions for mid-level women in technology*: Michelle R. Clayman Institute for Gender Research, Stanford University ....

Souitaris, V., Zerbinati, S., & Al-Laham, A. (2007). Do entrepreneurship programmes raise entrepreneurial intention of science and engineering students? The effect of learning, inspiration and resources. *Journal of Business Venturing, 22*(4), 566-591. doi:https://doi.org/10.1016/j.jbusvent.2006.05.002

Stewart, J., & Knowles, V. (2003). Mentoring in undergraduate business management programmes. *Journal of European Industrial Training*.

Stone, M. (1974). Cross-validatory choice and assessment of statistical predictions. *Journal of the Royal Statistical Society: Series B (Methodological), 36*(2), 111-133.

Sun, H., Lo, C. T., Liang, B., & Wong, Y. L. B. (2017). The impact of entrepreneurial education on entrepreneurial intention of engineering students in Hong Kong. *Management Decision, 55*(7), 1371-1393.

Teague, J. (2002). Women in computing: What brings them to it, what keeps them in it? *ACM SIGCSE Bulletin, 34*(2), 147-158.

Tenenhaus, M., Vinzi, V. E., Chatelin, Y.-M., & Lauro, C. (2005). PLS path modeling. *Computational Statistics & Data Analysis, 48*(1), 159-205. doi:https://doi.org/10.1016/j.csda.2004.03.005

Testa, S., & Frascheri, S. (2015). Learning by failing: What we can learn from un-successful entrepreneurship education. *The International Journal of Management Education, 13*(1), 11-22.

Thrash, T. M., & Elliot, A. J. (2003). Inspiration as a psychological construct. *Journal of personality and social psychology, 84*(4), 871.

Union, O. E. (2019). *The missing entrepreneurs 2019: policies for inclusive entrepreneurship*: OECD Publishing.

Van Auken, H., Fry, F. L., & Stephens, P. (2006). The influence of role models on entrepreneurial intentions. *Journal of developmental Entrepreneurship, 11*(02), 157-167.

Van Auken, H., Stephens, P., Fry, F. L., & Silva, J. (2006). Role model influences on entrepreneurial intentions: A comparison between USA and Mexico. *The International Entrepreneurship and Management Journal, 2*(3), 325-336. doi:10.1007/s11365-006-0004-1

Van Praag, C. M., & Versloot, P. H. (2007). The economic benefits and costs of entrepreneurship: A review of the research. *Foundations and Trends® in Entrepreneurship, 4*(2), 65-154.

Videnovik, M., Zdravevski, E., Lameski, P., & Trajkovik, V. (2018). *The BBC Micro: bit in the International Classroom: Learning Experiences and First Impressions.* Paper presented at the 17th International Conference on Information Technology Based Higher Education and Training (ITHET), Olhao, Portugal.

Voyles, M., Haller, S. M., & Fossum, T. V. (2007). Teacher responses to student gender differences. *ACM SIGCSE Bulletin, 39*(3), 226-230.

Welter, F. (2011). Contextualizing entrepreneurship—conceptual challenges and ways forward. *Entrepreneurship Theory and Practice, 35*(1), 165-184.

Westhead, P., & Solesvik, M. Z. (2016). Entrepreneurship education and entrepreneurial intention: do female students benefit? *International small business journal, 34*(8), 979-1003.

Wilson, F., Kickul, J., & Marlino, D. (2007). Gender, entrepreneurial self–efficacy, and entrepreneurial career intentions: Implications for entrepreneurship education. *Entrepreneurship Theory and Practice, 31*(3), 387-406.

Wold, H. (1982). Soft modeling: the basic design and some extensions. *Systems under indirect observation, 2*, 343.

Zhang, F., Wei, L., Sun, H., & Tung, L. C. (2019). How entrepreneurial learning impacts one's intention towards entrepreneurship: A planned behavior approach. *Chinese Management Studies, 13*(1), 146-170.

Zhang, Y., Duysters, G., & Cloodt, M. (2014). The role of entrepreneurship education as a predictor of university students' entrepreneurial intention. *International Entrepreneurship and Management Journal, 10*(3), 623-641.

Zhao, X., Lynch Jr, J. G., & Chen, Q. (2010). Reconsidering Baron and Kenny: Myths and truths about mediation analysis. *Journal of consumer research, 37*(2), 197-206.
20